\documentclass[aps,prl,twocolumn,showpacs,showkeys,nofootinbib,groupedaddress]{revtex4-1}
\usepackage{epsfig}
\usepackage{amsmath}
\usepackage{amsfonts}
\usepackage{amssymb}
\usepackage{graphicx}
\usepackage{colordvi}
\usepackage{color}
\usepackage{dcolumn}
\usepackage{multirow}
\usepackage{hyperref}
\usepackage{epstopdf}
\usepackage{bm}
\usepackage{times}
\hypersetup{
  colorlinks=true,        
  linkcolor=blue,         
  citecolor=magenta,      
}

\begin{document}

\title{Modified gravity for surveys}
                             
\author{Luisa G. Jaime}
\email{luisa@nucleares.unam.mx}
\affiliation{Instituto de Ciencias Nucleares, Universidad Nacional Aut\'onoma de M\'exico, A.P. 70-543 CDMX 04510, M\'exico}

\author{Mariana Jaber}
\email{jaber@estudiantes.fisica.unam.mx}
\affiliation{Instituto de F\'isica, Universidad Nacional Aut\'onoma de M\'exico, A.P. 20-364 CDMX 01000, M\'exico}

\author{Celia Escamilla-Rivera}
\email{cescamilla@mctp.mx}
\affiliation{Mesoamerican Centre for Theoretical Physics, Universidad Aut\'onoma de Chiapas.
Ciudad Universitaria, Carretera Zapata Km. 4, Real del Bosque (Ter\'an), 29040, Tuxtla Guti\'errez, Chiapas, M\'exico.}

\date{\today}

\begin{abstract}
{We present a new parameterization for the equation of state (EoS) $\omega_X=P_X/\rho_X$, which can reproduce a $f(R)$-like evolution with a precision between $[0.5\%-0.8\%]$ over the numerical solutions. Also, our proposal can render a variety of popular $f(R)$ models that are considered as viable candidates for the cosmic late time acceleration. By using observational data from baryonic acoustic oscillations, supernovae and cosmic chronometers we investigate the constraints on the new EoS parameters. This proposal set a EoS formulation which can be used in an efficient way and makes a good candidate to be implemented in a variety of surveys in order to test the $f(R)$ generic behaviour.}
\end{abstract}

\pacs{
98.80.−k  
04.50.Kd, 
95.36.+x  
95.80.+p  
}
\maketitle


\section{Introduction}
Currently, the Universe go through an accelerated expansion, different observations have proved such fact: Supernovae Type Ia (SNIa) \cite{Riess:1998cb-Perlmutter:1998np}, Baryon Acoustic Oscillations (BAO) \cite{Eisenstein:2005su}, Cosmic Microwave Background Radiation (CMBR) anisotropies \cite{Spergel:2003cb}, Large Scale Structure formation \cite{Tegmark:2003ud} and Weak Lensing \cite{Jain:2003tba}. Even more, future projects and surveys \cite{surveys} are underway or being proposed to discover the underlying cause of this phenomena. 

The current standard cosmological paradigm 
is 
the $\Lambda$CDM model, where $\Lambda$ is a constant term added to the Einstein-Hilbert action. Such addition is performed  in order to produce late time acceleration and even when there is no physical ground to justify $\Lambda$, making this model able to fit all the current observations. However, some degrees of tension have appeared among different data sets which has renewed the interest on alternative models that can provide an acceleration mechanism.
For instance, the value for the matter density fraction consistent with the Lyman-$\alpha$ forest measurement of the baryon acoustic oscillations \cite{Delubac:2014aqe} is smaller than the one preferred by CMB measurements. Also, the value of $H_0$ inferred from the Planck CMB data \cite{Ade:2015xua} is $3.4$ $\sigma$ lower than the local measurement reported by \cite{Riess:2016jrr}. 
A promising new and independent measurement of $H_0$ will come from standard-siren measurements from  gravitational waves sources and provide more information on the issue, although the tension linger with the current measurement  \cite{Abbott:2017xzu}.

Using all current observational data, in \cite{Zhao:2017cud} was reconstructed the Dark Energy equation of state (EoS) obtaining a distinctive shape that crosses multiple times the phantom divide line. This kind of oscillating $\omega_X$ can not be produced with a single phantom or quintessence field \cite{Shafieloo:2012rs} but it can be produced by modified gravity. 

Hereafter, we will focus our attention in $f(R)$ gravity models. Their characteristic EoS \cite{Jaime:2013zwa} makes them an appealing framework in order to reproduce the dynamical evolution of $\omega_X$ found in \cite{Zhao:2017cud}. In regards to the tension issue, in 
modified gravity theories (e.g. Galileon) may reconcile the Planck with high $H_0$ values \cite{Barreira:2014jha,Escamilla-Rivera:2015ova}, although known models have problems with either cosmology \cite{Renk:2017rzu} or gravitational waves \cite{Ezquiaga:2017ekz}. In this context, the resulting field equations are of fourth order on the metric and behave like attractors; therefore their implementation in the pipeline of surveys, or in N-body, and Boltzmann codes requires many assumptions.

In this paper we present the construction of a new parameterization for the EoS in order to reproduce a variety of $f(R)$ models between $[0.5\%-0.8\%]$ of precision which can help to test these models in a straightforward way.
This parameterization can be used as a fiducial model in surveys with the advantage that this one has a physical motivation in comparison to some others like CPL \cite{Chevallier:Polarski,Linder}.
Future surveys like Euclid \cite{2011arXiv1110.3193L} and DESI \cite{Levi:2013gra, Aghamousa:2016zmz} will play a fundamental role in the understanding cosmic acceleration and will allow us to test interesting models of gravity and dark energy.


\section{$f(R)$ cosmology, equation of state and models.}
\label{sec:fR}
These theories of gravity take a general function of the Ricci scalar in the Einstein-Hilbert action
\begin{equation}
\label{f(R)}
S[g_{ab},{\mbox{\boldmath{$\psi$}}}] =
\!\! \int \!\! \frac{f(R)}{2\kappa} \sqrt{-g} \: d^4 x 
+ S_{\rm matt}[g_{ab}, {\mbox{\boldmath{$\psi$}}}] \; ,
\end{equation}
where $G=c=1$ and $\kappa \equiv 8\pi$, $S_{\rm matt}[g_{ab}, {\mbox{\boldmath{$\psi$}}}]$ is the usual 
action for matter. $f(R)$ is an arbitrary smooth function of the Ricci scalar $R$. The field equations, associated to this action, are given by:
\begin{equation}
\label{fieldeq1}
f_R R_{ab} -\frac{1}{2}fg_{ab} - 
\left(\nabla_a \nabla_b - g_{ab}\Box\right)f_R= \kappa T_{ab}\,\,,
\end{equation}
where $f_R = \partial_R f$, $\Box= g^{ab}\nabla_a\nabla_b$ and $T_{ab}$ is the energy-momentum tensor for matter. They can be rewritten as:


\begin{eqnarray}
\label{fieldeq3}
& G_{ab} =& \frac{1}{f_R}\Bigl{[} f_{RR} \nabla_a \nabla_b R +
 f_{RRR} (\nabla_aR)(\nabla_b R) \nonumber \\
&  & -\frac{g_{ab}}{6}\Big{(} Rf_R+ f + 2\kappa T \Big{)} 
+ \kappa T_{ab} \Bigl{]}, \; 
\end{eqnarray}

where $G_{ab}= R_{ab}-g_{ab}R/2$ is the Einstein tensor. In the present work we are using the Ricci scalar approach to $f(R)$ proposed in \cite{Jaime:2010kn} and then used in cosmology (\cite{Jaime:2015afa,Berti:2015itd}). 

We will consider a homogeneous, isotropic universe described by the Friedman-Lema\^itre-Robertson-Walker (FLRW) metric:

\begin{equation}
\label{SSmetric}
ds^2 = - dt^2  + a^2(t)\!\left[ \frac{dr^2}{1-k r^2} + r^2 \left(d\theta^2 + \sin^2\theta d\varphi^2\right)\right]
\!,\!
\end{equation}
with $k=0$. The energy momentum tensor (EMT) is that for a fluid composed by baryons, dark matter and radiation. Under these assumptions, we will obtain a second order differential equation for the Ricci scalar by taking the trace of (\ref{fieldeq1}) and the modified Friedman equations from (\ref{fieldeq3}).

\begin{eqnarray}
\label{traceRt}
& \ddot R = &-3H \dot R -  \frac{1}{3 f_{RR}}\left[ 3f_{RRR} \dot R^2 + 2f- f_R R + \kappa T \right], \,\,\,\,\,\, \\
\label{Hgen}
& H^2 = & -\frac{1}{f_{RR}}\left[f_{RR}H\dot{R}-\frac{1}{6}(Rf_{R}-f) \right]-\frac{\kappa T^{t}_{t}}{3f_{R}}, \\
\label{Hdotgen}
& \dot{H}= & -H^2 -\frac{1}{f_{R}} \left[ f_{RR}H\dot{R} + \frac{f}{6}+\frac{\kappa T^{t}_{t}}{3} \right]  \,\,\,,
\end{eqnarray}
where $H = \dot a/a$. The EoS~\footnote{This choice is obtained by defining $T_{ab}^X=T_{ab}^{tot}-T_{ab}$ where $T_{ab}^X$ is the energy momentum tensor (EMT) associate with the geometric dark energy in $f(R)$, $T_{ab}^{tot}$ is the total EMT and $T_{ab}$ is the EMT associated to the matter lagrangian. This choice of the EoS has no degeneracies (see \cite{Jaime:2012gc} for a discussion about the EoS in $f(R)$)} for the geometric dark energy in $f(R)$
is given by:
\begin{equation}
\label{eq:GEoS}
\omega_X = \frac{3H^2-3\kappa P-R}{3(3H^2-\kappa\rho)},
\end{equation}
where the Ricci scalar is given by $R = 6(\dot{H}+2H)$, $P$ and $\rho$ are presure and density, respectively, of the matter and radiation content. The models used in this work can provide an accelerated evolution, with a $\omega_X \approx -1$. In the case of (\ref{eq:f(R)-HS}) and (\ref{eq:f(R)-St}) such evolution goes asymptotically to the de Sitter point ($R_{(z \rightarrow -1)} > 0$). In the case of (\ref{eq:f(R)-exp}) the future is asymptotically $R_{(z \rightarrow -1)} = 0$ with a transient but apparently long enough accelerated epoch.

Some of the most successful $f(R)$ models in cosmology are: 
\begin{itemize}
 \item[a) ] Hu $\&$ Sawicki model~\footnote{In the Hu-Sawicki model the parameters $c_1$ and $c_2$ are related with $f_R^0$ and $\Omega_M^0$ according as is explained in \cite{Hu:2007nk}} \cite{Hu:2007nk}
   			 \begin{equation}\label{eq:f(R)-HS}
   				 f(R)= R- R_{\rm HS}\frac{c_{1}\left(\frac{R}{R_{\rm HS}}\right)^n}{c_{2}\left(\frac{R}{R_{\rm HS}}\right)^n+1} \; ,
    		 \end{equation}
 \item[b) ] Starobinsky model \cite{Starobinsky:2007hu}
          	 \begin{equation}\label{eq:f(R)-St}
          	 f(R)= R+\lambda R_{S}\left[ \left( 1+\frac{R^2}{R^2_{S}}\right)^{-q}-1\right],
			 \end{equation}     
 \item[c) ] The exponential model \cite{EXP} 
             \begin{equation}\label{eq:f(R)-exp}
             f(R)=R+\beta R_* (1-e^{-R/R_*}).
             \end{equation}                  	  
\end{itemize}
All the parameters involved in these functions should be constrained according to observations. In order to perform such tests we need to integrate the field equations from the past to the future and, given the attractor behaviour of this kind of gravity, its implementation into Boltzmann codes for alternative models \cite{Zumalacarregui:2016pph} or surveys is complex.


\section{Parametric EOS for $f(R)$.}

Parameterizations of the EoS, for the accelerating mechanism in the universe containing two \cite{Escamilla-Rivera:2016qwv} or more parameters have been proposed in the literature, either inspired by the behaviour of scalar-field dynamics \cite{Jaber:2017bpx} or motivated by the tomographic reconstruction of BAO data \cite{Wang:2016wjr}. As we mentioned, the results shown in \cite{Zhao:2017cud} leads to an Universe with a dynamical and no monotonic dark energy. If this result prevails using future surveys, such dynamics will involve an oscillating EoS for the accelerating mechanism. 


\begin{table}
\begin{center}
\begin{tabular} { llcclll }
\hline
$f(R)$ model       & \,\,\,\, &$\Omega_M(z=0)$ \,\,\, & Parameter values &  \\
\hline
   			       & \,\,\,\, & $0.20$            &$c_2=2.78\times10^{-5}$ & \\    
Hu-Sawicki         & \,\,\,\, & $0.25$            &$c_2=7.98\times10^{-5}$ & \\    
                   & \,\,\,\, & $0.30$            &$c_2=1.95\times10^{-4}$ & \\    
\hline
 		           & \,\,\,\, & $0.20$            &$\lambda=1.15$, $R_S=1$ & \\
Starobinsky $2007$ & \,\,\,\, & $0.25$			  &$\lambda=1.0$,   $R_S=1$ & \\
			       & \,\,\,\, & $0.30$      	  &$\lambda=0.9$,   $R_S=1$ & \\
\hline
                   & \,\,\,\, & $0.20$      	  &$\beta=0.5$, $R_*=5$  & \\
Exponential        & \,\,\,\, & $0.25$        	  &$\beta=0.8$, $R_*=5$  & \\
                   & \,\,\,\, & $0.30$      	  &$\beta=0.6$, $R_*=6$  & \\
\hline
\end{tabular}
\caption{First column: $f(R)$ model, second column: $\Omega_{m}^{0}$ and third column value of the parameters for each model. For the Hu-Sawicki model we have computed, for the three cases, the values for $c_2$ by taking $R_{HS}=1$ and $f_{R}^{0}=0.01$ and the corresponding $\Omega_m^0$ value, $c_1$ will be given by $c_1= c_{2}6(1-\Omega_m^0)/\Omega_m^0$ (see \cite{Hu:2007nk} for a detailed explanation). For the Starobinsky model we have taken $n=2$ for the three cases.}
\label{Table:Parameters}
\end{center}
\end{table}

In order to provide a useful way to implement a $f(R)$-like cosmology in observational tests, surveys or Boltzmann codes, we build a new parameterization 
involving four parameters. This parameterization is based on the numerical results coming from the integration of the field equations in $f(R)$.

The numerical integration is performed by using a fourth order Runge-Kutta integrator. Initial conditions are fixed in the past at some value of $z$ where $\Omega_M(z)$ is very close to $1$, the value of the EoS for the geometric dark energy is $\omega_X=-1$ at this value of redshift. The Hamiltonian constrains imposed by $H^2$ in (\ref{Hgen}) is used as an internal test in the code (see~\cite{Jaime:2012gc} for a detailed revision about the implementation in cosmology). We perform the numerical integration for the three models (\ref{eq:f(R)-HS}$)-($\ref{eq:f(R)-exp}) presented in the previous section. The values for the parameters of each $f(R)$ model are listed in table (~\ref{Table:Parameters}) as well as the value of $\Omega_M(z=0)$. 

By integrating the filed equations we will obtain the evolution for $R$ and $H$ and also $\dot{H}$, this is the information we need to compute the equation of state $\omega_X$ given by (\ref{eq:GEoS}). According to this, our proposal for a parametric EoS in $f(R)$ is given by the following function

\begin{equation}
\label{eq:fit}
\omega(z) = -1 + \frac{w_0}{1+w_1z^{w_2}}cos(w_3+z).
\end{equation}
where $\omega_i$ are free parameters and $z$ is the standard redshift given by $z=a_0/a-1$. We notice that (\ref{eq:fit}) has a present value given by $w(z=0) = w_0cos(w_3)-1$, recovers $\omega_X=-1$ at large redshifts and allows oscillations in the range of interest for observations and future surveys. We use Mathematica software in order to fit the cosmological parameters. It should be noted that this ensamble use the Levenberg–Marquardt algorithm by default, but also allows to choose among several other algorithms for function minimization, which in our case we got a fit precision of $10^{-10}$.

\begin{figure}
\includegraphics[scale=0.30]{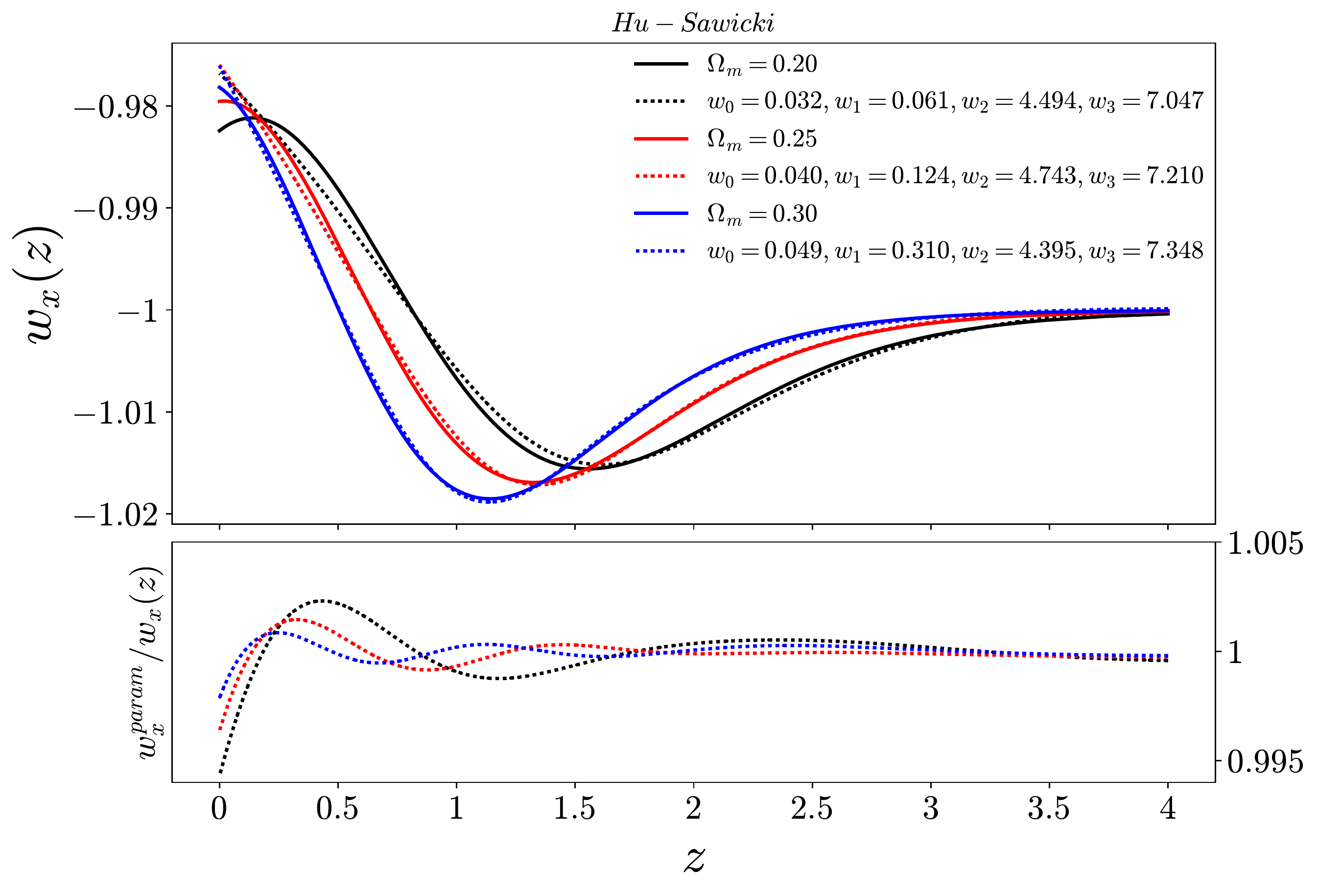}
\caption{Upper frame: Geometric dark energy equation of state for the Hu-Sawicki model (\ref{eq:f(R)-HS}) with different values of $\Omega_m^0$. Solid lines represent the numerical integration of the field equations and their reconstruction (in dashed lines) comes from the best fit by using \eqref{eq:fit} (JJE parameterization). a) Black is for $\Omega_{m}^{0}=0.20$, b) Red is for  $\Omega_{m}^{0}=0.25$ and c) Blue is for  $\Omega_{m}^{0}=0.30$.  Best fit parameters are shown in the plot for each case. Bottom frame: Evolution of the ratio parameterized $\omega_{X,param}$ and numerical EoS $\omega_X$, values remains within $ 0.5$\%.}
\label{fig:Fit-HS}
\end{figure}

\begin{figure}
\includegraphics[scale=0.30]{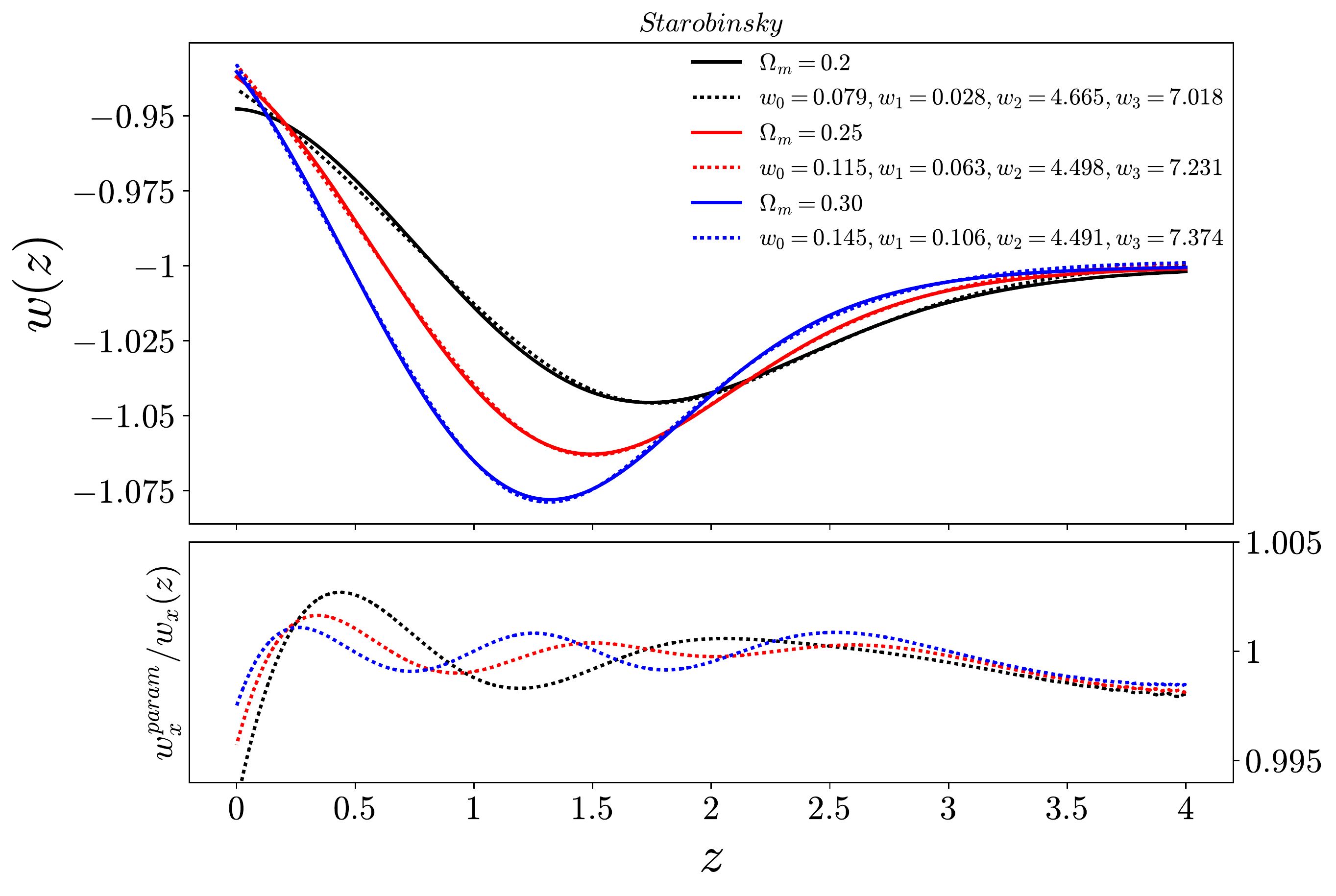}
\caption{Upper frame: Same as figure [\ref{fig:Fit-HS}] for the Starobinsky model (\ref{eq:f(R)-St}). Bottom frame: Same as bottom frame in [\ref{fig:Fit-HS}]}
\label{fig:Fit-St}
\end{figure}

\begin{figure}
\includegraphics[scale=0.30]{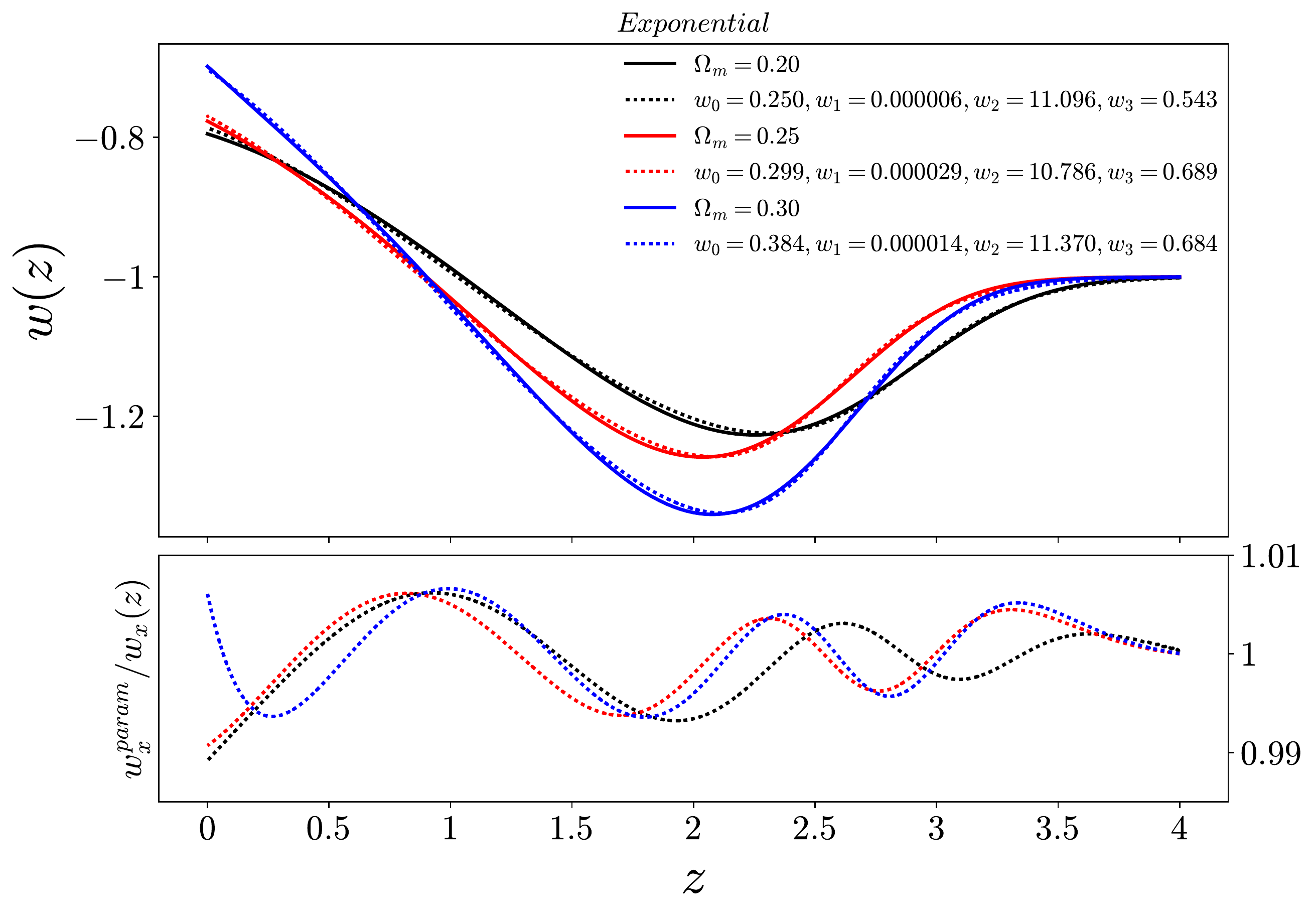}
\caption{Upper frame: Same as figure [\ref{fig:Fit-HS}] for the Exponential model (\ref{eq:f(R)-exp}). Bottom frame: Same as bottom frame in [\ref{fig:Fit-HS}] but, in this case, best fits are within $ 1$\%}
\label{fig:Fit-Exp}
\end{figure}

Figures [\ref{fig:Fit-HS}-\ref{fig:Fit-Exp}] shows the evolution for the models  (\ref{eq:f(R)-HS}$)-($\ref{eq:f(R)-exp}) and the best fit for each one of them by using our proposal (\ref{eq:fit}). The evolution can be recovered for (\ref{eq:f(R)-HS}) and (\ref{eq:f(R)-St}) within a $0.5\%$ while for (\ref{eq:f(R)-exp}) fits are within a $0.8\%$ precision. These are reasonable values where current and future experiments can set a cut off over the cosmological parameters, e.g. for BOSS (BGS) and BOSS \cite{DESIref} we have an enough statistical significance for the JJE parameterisation at $1\%$ below $z=1$. Between $z=1$ and $z=2$, eBOSS and EUCLID would be within 1\% accuracy for JJE \cite{Aghamousa:2016zmz}.

\section{JJE implementation to observational data.}

Given that we are interested in modelling the late-time evolution of the universe we use observational data from BAO redshift surveys, SNeIA luminous distance from  Union 2.1 \cite{union21} and the latest high-z measurements of $H(z)$ from Cosmic Chronometers \cite{cc}.
 
We use  measurements of the BAO peak from the galaxy redshift surveys six-degree-field galaxy survey (6dFGS \cite{Beutler:2011hx}), 
Sloan Digital Sky Survey Data Release 7 (SDSS DR7 \cite{Ross:2014qpa}) and the reconstructed value (SDSS(R)  \cite{Padmanabhan2pc}), as well as the latest result from the complete BOSS sample SDSS DR12 (\cite{Alam:2016hwk}), and also from the Lyman-$\alpha$ Forest  measurements from the Baryon Oscillation Spectroscopic Data Release 11 (BOSS DR11 \cite{Font-Ribera:2013wce}, \cite{Delubac:2014aqe}).  
Since the volume surveyed by BOSS and WiggleZ \cite{Kazin:2014qga} partially overlap we do not use data from the latter in this work (see details in \cite{Beutler:2015tla}).
Even though the current supernovae compilation is given by the JLA sample \cite{Betoule:2014frx}, in this work we implement the Union 2.1 sample since the apparent magnitude ratio is less than $0.2\%$ in the redshift range of our interest (above $z=1$) in comparison to the JLA sample.

In addition to the free parameters in \eqref{eq:fit} we vary the fractional amount of matter $\Omega_M$ and the value of $H_0$, by means of a standard $\chi^2$ approach we find the constraints at 1 and 2-$\sigma$ level.

\begin{figure}
	\centering
	\includegraphics[width=0.9\linewidth]{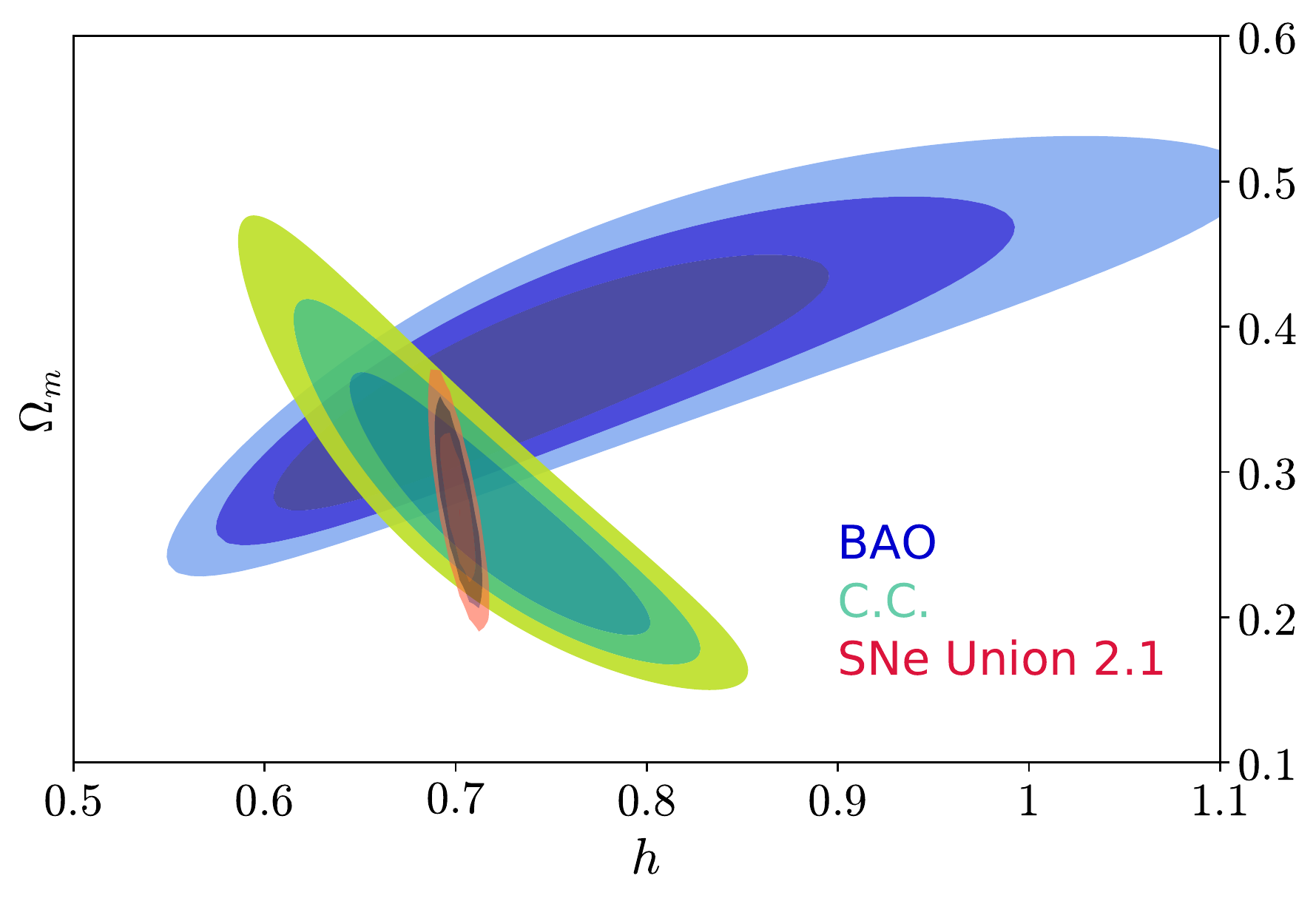}
	\caption{Constraints on the $\Omega_M$-$h$ space for the JJE parametrization \eqref{eq:fit}. }
	\label{fig:fig-frcontours}
\end{figure}

\begin{figure}
	\centering
\includegraphics[width=0.9\linewidth]{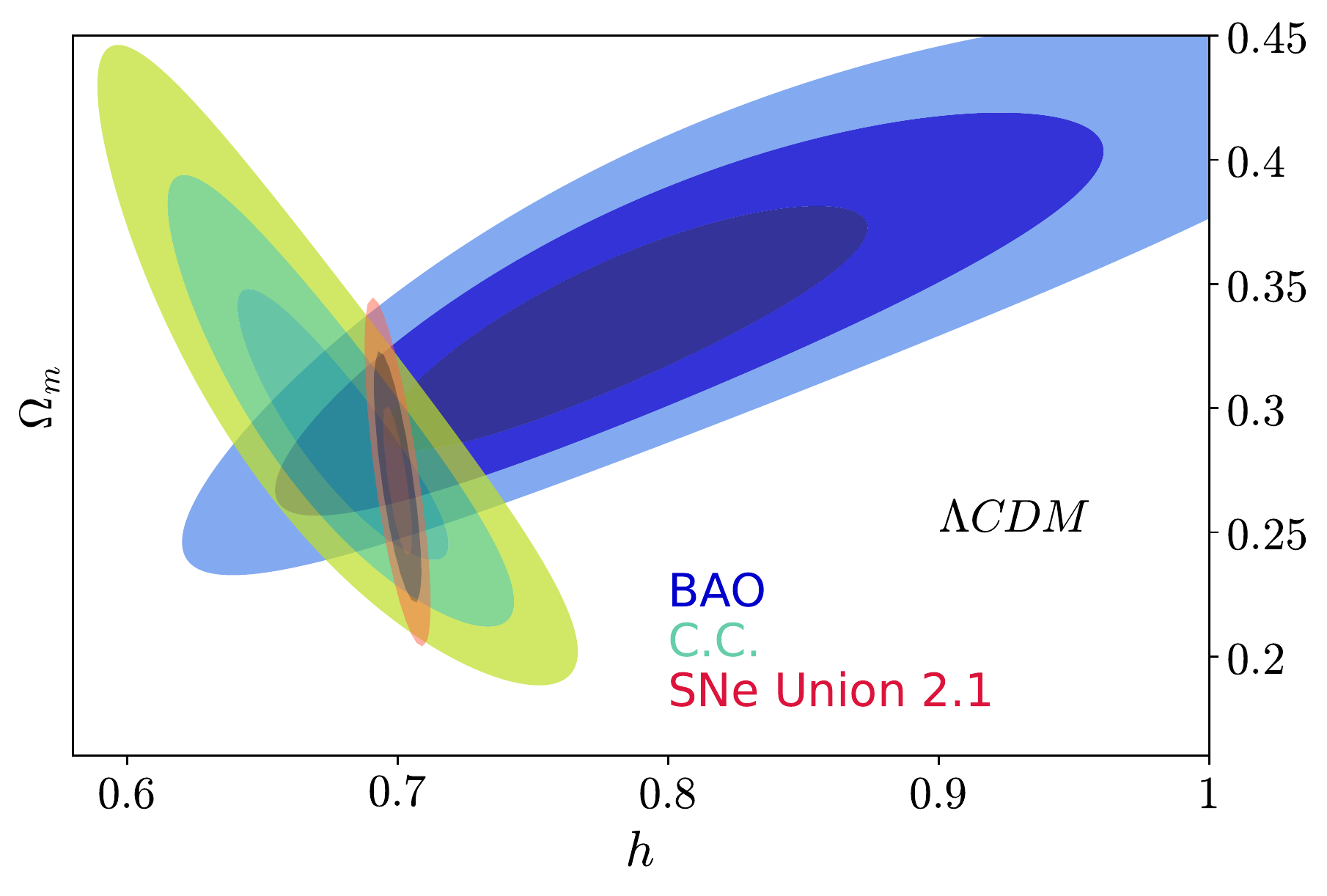}	\caption{Constraints on the $\Omega_M$-$h$ space for  $\Lambda$CDM model.}
	\label{fig:fig-lcdmcontours}
\end{figure}

In Figures (\ref{fig:fig-frcontours}) and (\ref{fig:fig-lcdmcontours}) we notice that the constraints on  \\ $\Omega_M$-$h$ parameter space are tighter for $\Lambda$CDM scenario in comparison with the JJE parameterisation. As for the contours at 1-$\sigma$ from the different datasets it is noticeable that they do overlap for the JJE parameterisation while a tension with
${\Delta \chi^2}_{\text{reduced (SN-BAO)}} =0.961$ and ${\Delta \chi^2}_{\text{reduced (CC-BAO)}} =0.348$ is present among the BAO  and supernovae results when $\Lambda$CDM is assumed. 

To compare our JJE model with $\Lambda$CDM we use the combination of the three different datasets (SN+BAO+CC) and calculate the corresponding reduced-$\chi^2$ estimator by taking into account the different number of degrees of freedom among the two models.
From the obtained  values we find that JJE parameterisation and $\Lambda$CDM  are consistent showing a difference of $\Delta\chi^2_{\text{reduced}(\text{JJE}-\Lambda \text{CDM})}=0.5\%$.


\section{Discussion.}
The scientific community is devoting a large amount of time and resources in the quest to understand the dynamics and nature of dark energy, working on current (SDSS-IV \cite{Dawson:2015wdb}, DES \cite{Abbott:2005bi}) and future  (DESI \cite{Levi:2013gra, Aghamousa:2016zmz}, Euclid \cite{2011arXiv1110.3193L}, LSST \cite{2009arXiv0912.0201}) experiments to study with very high precision the expansion history of the universe and thus be able to test interesting theoretical models. In the process of analyzing data coming, for instance, from galaxy redshift surveys, a cosmological model is used throughout the pipeline 
(\cite{RSD-paper}). Also, analysis of the CPL parameterisation using forecast for the eBOSS has been done in \cite{Ruggeri:2017rza} to convert observed positions   of the objects into coordinates. 

Therefore, to implement in an easy and efficient way modified gravity theories in any kind of survey, we proposed the JJE parameterisation (\ref{eq:fit}).
Similarly, in future forecast analysis (\cite{DESI-forecasting}) a cosmological model will be needed to investigate the 
parameter constraints in modified gravity theories.

With the presented proposal we aim to put theoretical background to parameterizations of $\omega_X$ and also models for $f(R)$ gravity at the same level as other parameterisations into the pipeline and analysis of observational data and forecasts.

It is worth to mention that by introducing this parameterisation in surveys or using it for data analysis we are avoiding any other kind of assumption that are usually taken in $f(R)$. One of the most usual assumptions is the one related to the value of $f_R(z=0)$~\footnote{Some other authors (like \cite{Hu:2007nk}) use a different notation, $f(R)=R+f(R)_{others}$ and constrictions to our $f_R$ are given by $f_R=1+f_{others}$}  which is taken very close to $1$ because of the Solar System constrains~\cite{Hu:2007nk} or the structure formation~\cite{EXP}. Nevertheless it is important to notice this constrictions are usually computed for the Hu-Sawicki model and such values do not necessarily apply to other models. By using the JJE parametrization we are making no assumption whatsoever over such values. 


\textit{Acknowledgements.-} The Authors thanks to M. Zumalacarregui for taking the time of reading these ideas and give us fruitful feedback. C.E-R. acknowledges MCTP-UNACH. M.J. thanks to CONACyT for the PhD fellowship.


\end{document}